\documentclass[prl, twocolumn, floatfix ]{revtex4}
\usepackage[dvips]{graphicx}
\usepackage{bm}

\begin{document}

\title{Infrared Hall effect in underdoped and optimally doped LSCO }
\author{L. Shi}
\author{D. Schmadel}
\author{H. D. Drew}
\affiliation{Department of Physics, University of Maryland,
College park, Maryland 20742}
\author{I. Tsukada}
\author{Yoichi Ando}
\affiliation{Materials Science Research Laboratory, Central
Research Institute of Electric Power Industry, 2-11-1 Iwadokita,
Komae, Tokyo 201-8511, Japan}

\begin{abstract}

We report a study of magneto-optical properties in the
mid-infrared region of a series of La$_{2-x}$Sr$_x$CuO$_4$ samples
with hole doping level ranging from severely underdoped
($x=0.03$)to optimally doped ($x=0.15$). The Faraday rotation and
circular dichroism are measured in a magnetic field of 8 Tesla and
in a temperature range between 30K and 300K.  The doping and
temperature dependence of infrared Hall angle is found to be
understood within a simple Drude model.  A significant increase of
Hall frequency is observed when the hole doping level is reduced
from optimal doping, which is compared with models of the
pseudogap.

\end{abstract}

\maketitle

%\small
As the hole doping in cuprate superconductors is reduced below
optimal $T_c$ they enter a pseudogap phase before they ultimately
undergo a transition to a Mott insulating ground state. While the
pseudogap phase exhibits evidence for a partial gapping of the
Fermi surface in many expereiments its character is otherwise
poorly understood. It is this mysterious pseudogap phase that
attracts the greatest attention in the quest for understanding the
mechanism for high $T_c$.  Many different scenarios have been
proposed for this state~\cite{orenstein01} . One of the proposals
for partially gapping the Fermi surface is the formation of a
density wave state. In this scenerio the large Fermi surface of
the optimally doped system develops energy gaps at the magnetic
Brillioun zone boundaries and breaks up into small pockets.
However no evidence of energy gaps have been reported from IR
studies and the evidence from ARPES is controversial. Recently,
Rigal \textsl{et.al.} \cite{rigal01} observed a dramatic increase
of the Hall frequency in underdoped YBa$_2$Cu$_3$O$_{6+x}$ (YBCO)
samples, which is consistent with the presence of Fermi pockets
due to a partial gapping of the original Fermi surface. An
alternative interpretation of the increase in the Hall frequency
follows from recent work by H. Kontani which includes vertex
corrections in the calculation of the optical conductivity of the
cuprates based on the fluctuation exchange model of the
interactions~\cite{kontani01,kontani02}.  In any case the analysis
of the results in YBCO is complicated by the existence of CuO
chains. Therefore it is highly desirable to repeat these
magneto-optical studies with other cuprates.

La$_{2-x}$Sr$_x$CuO$_4$ (LSCO) is an ideal compound to study. It
has a relatively simple lattice with one  layer of CuO$_2$ plane,
and without the complication of CuO chain as in YBCO.  And a wide
range of doping level is readily available.  Of particular
interest is the severely underdoped LSCO, which is predicted by
conventional phase diagram of cuprates to be anti-ferromagnetic
insulator. However recently angle-resolved photoemission
(ARPES)~\cite{yoshida01}, DC transport~\cite{ando01} and Hall
effect~\cite{ando02} measurements show evidences of metallic
behavior even for $x=0.02$ sample. Furthermore, measurement of
anisotropy of DC~\cite{ando03} and optical~\cite{dumm01}
conductivity of de-twinned samples indicates possible existence of
charge stripes. In this letter we report the study of
magneto-optical properties in the mid-infrared region of a series
of LSCO samples ranging from slightly hole doped to optimally
doped, by measuring the Faraday rotation and circular dichroism in
a magnetic field of 8 Tesla and in a temperature range between 30K
and 300K. We found that the doping and temperature dependence of
Hall angle can be understood well understood with a simple Drude
model.  A significant increase of Hall frequency is observed when
the hole doping level is reduced from optimal doping,  which is
consistent with drastic reduction of the volume of Fermi surface
in the underdoped sample.

The La$_{2-x}$Sr$_x$CuO$_4$ samples used in this study are
severely underdoped ($x=0.03$), underdoped ($x=0.10$) and
optimally doped ($x=0.15$) thin films grown on SrTiO$_3$
substrates by pulsed laser deposition (PLD).  The $x=0.03$ samples
does not show superconducting transition, while the other two
samples have $T_c$ of 32K and 35K, respectively. The thin films
have thickness ranging from 2500 $\mathrm{{\AA}}$ to 3800
$\mathrm{{\AA}}$.

In the infrared Hall study various magneto-optical properties such
as Hall angle $\theta^{}_{\mathrm{H}}$ and Hall conductivity
$\sigma_{xy}$ are obtained by measuring the complex Faraday angle,
whose real and imaginary parts  correspond  to Faraday rotation
and circular dichroism when the linearly polarized laser beam from
a CO$_2$ laser pass through the sample. In the mid-infrared region
(900cm$^{-1} \sim$ 1100cm$^{-1}$) where the measurements are
taken, the typical value of Faraday angle is at the order
$10^{-4}\sim 10^{-5}$ radians per Tesla.  Such a precision
measurement of   Faraday angle requires a very sensitive technique
which is achieved  by using a ZnSe photoelastic modulator to
analyze the change in the polarization of the laser
beam~\cite{cerne02}.

\begin{figure}[tbh]
  \includegraphics[width=8.3cm]{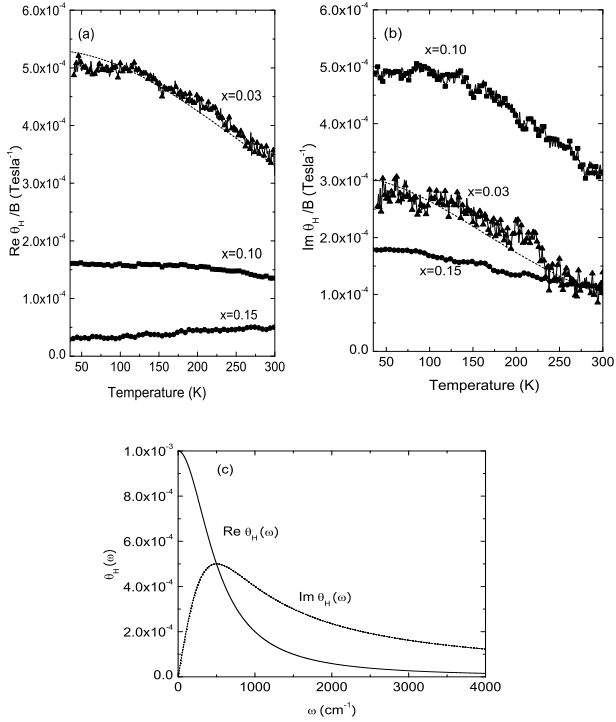}\\
  \caption{ (a) and (b):  Temperature dependence of real
  (a) and imaginary (b) parts of Hall angle $\theta^{}_{\mathrm{H}}$ at $\omega = 1087
  \mathrm{cm}^{-1}$ for x=0.03, 010 and 0.15 samples. (c): Simulated
  frequency dependence of Hall angle based on simple Drude model
  with $\gamma^{}_{\mathrm{H}}= 500 \mathrm{cm}^{-1}, \omega^{}_{\mathrm{H}} = 0.5$.
    }
  \label{fg:001}
\end{figure}

It has been proved that the optical Hall angle $\tan
\theta^{}_{\mathrm{H}}(\omega)\sim \theta^{}_{\mathrm{H}}(\omega)
= \sigma_{xy}(\omega)/\sigma(\omega) $ is a response function with
it own $f$-sum rule~\cite{drew01}.   Therefore similar to other
response function like optical conductivity, the simplest
functional form of Hall angle is that of the Drude model:
\begin{equation}\label{eq:001}
    \theta^{}_{\mathrm{H}}(\omega) = \frac{\omega^{}_{\mathrm{H}}}
    {\gamma^{}_{\mathrm{H}}-i \omega},
\end{equation}
where the Hall scattering rate $\gamma^{}_{\mathrm{H}}$ is a
different integral~\cite{cerne01} of $\gamma(\bm{k})$ along the
Fermi surface than that of the regular scattering rate, while the
Hall frequency $\omega^{}_{\mathrm{H}}$ is related to the
cyclotron frequency in the simple metal.  Study in past few years
suggests that the Drude model of Hall angle works well for simple
metal films like Au and Cu, as well as hole doped cuprates with
various doping, except for electron doped Pr$_{2-x}$Ce$_x$CuO$_4$,
in which there is evidence of the formation of a spin-density wave
state   in the underdoped region.
%Recently,
%Kontani~\cite{kontani01, kontani02} studied infrared Hall effect
%in cuprates based on fluctuation-exchange (FLEX)~\cite{yanase01}
%approximation and full current-vertex correction (CVC).  His
%calculation reproduces the simple Drude-form for the Hall angle
%and shows that the extended Drude breaks down for Hall
%conductivity.

Shown in figure~\ref{fg:001}(a) and (b) are the temperature
dependence of real(a) and imaginary(b) parts of the complex Hall
angles $\theta^{}_{\mathrm{H}}$,  measured at $\omega
=1087$cm$^{-1}$, for $x=0.03$, $x=0.10$ and $x=0.15$ samples,
respectively. When the hole doping level is increased from the
severely underdoping level at $x=0.03$ to the optimally doped
level at $x=0.15$, the real part of $\theta^{}_{\mathrm{H}}$
decrease monotonically with increasing doping, while the imaginary
part first increases, then decreases with increasing doping.  The
observed doping dependence of $\theta^{}_{\mathrm{H}}$ can be
understood if one consult figure~\ref{fg:001}(c), which is a plot
of $\theta^{}_{\mathrm{H}}(\omega)$ based on the simple Drude
model (equation \ref{eq:001}).   Similar to other more familiar
response function like optical conductivity, the real part of
$\theta^{}_{\mathrm{H}}(\omega)$ is a Lorentzian centered at
$\omega=0$ with half width $\gamma^{}_{\mathrm{H}}$, while the
imaginary part reaches its peak at $\gamma^{}_{\mathrm{H}}$.
Therefore the doping dependence of $\theta^{}_{\mathrm{H}}$
  can be understood if we assume that
$\gamma^{}_{\mathrm{H}}$ of the $x=0.03$ sample is larger than
$\omega=1087$cm$^{-1}$ and the other two samples have
$\gamma^{}_{\mathrm{H}}$ smaller than $\omega $.

\begin{figure}[tbh]
  % Requires \usepackage{graphicx}
  \includegraphics[width=8.3cm]{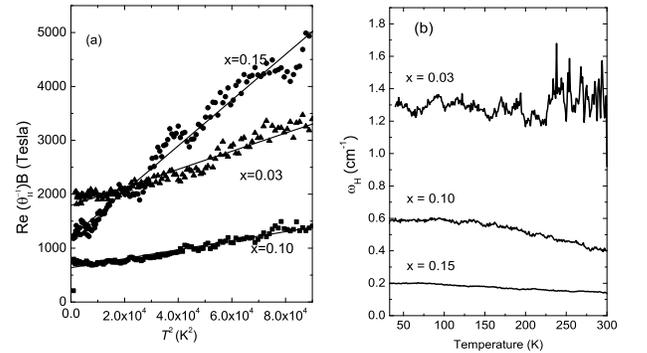}\\
  \caption{ (a) Real part of inverse Hall angle $\theta^{-1}_{\mathrm{H}}$ measured at $\omega = 1087
  \mathrm{cm}^{-1}$ for $x=0.03, 0.10$ and 0.15 samples.
  (b): Hall frequencies $\omega^{}_{\mathrm{H}}$ of the x=0.03,
  0.10 and 0.15 samples.
   }
\label{fg:003}
\end{figure}

Plotted in figure~\ref{fg:003}(a) are the real parts of inverse
Hall angles as functions of square of temperature for the three
samples, in temperature range from 30K to 300K,  as well as their
linear fits.  Since $\theta^{-1}_{\mathrm{H}} \sim \cot
\theta^{}_{\mathrm{H}}  $ for small $\theta^{}_{\mathrm{H}}$ ,  it
indicates that $\cot \theta^{}_{\mathrm{H}} \propto T^2 $ for all
three samples, which is consistent with recent result from DC Hall
effect~\cite{ando02}.
%, but in contrast to $\cot
%\theta^{}_{\mathrm{H}} \propto T  $ for  other underdoped cuprates
%(ref?).
The $T^2$ linear dependence of $\cot \theta^{}_{\mathrm{H}}$, and
therefore the Hall scattering rate $\gamma^{}_{\mathrm{H}}$ can be
demonstrated by fitting the temperature dependence of
$\theta^{}_{\mathrm{H}}$ by the simple Drude model:
\begin{equation}\label{eq:002}
    \mathrm{Re}\ \theta^{}_{\mathrm{H}}(T) = \frac{\omega^{}_{\mathrm{H}}
    \gamma^{}_{\mathrm{H}}(T)}
    {\gamma^{2}_{\mathrm{H}}(T)+\omega^2}, \quad
    \mathrm{Im}\ \theta^{}_{\mathrm{H}}(T) = \frac{\omega^{}_{\mathrm{H}}
    \omega}
    {\gamma^{2}_{\mathrm{H}}(T)+\omega^2},
\end{equation}
with a Hall scattering rate linearly dependent on $T^2$:
$\gamma^{}_{\mathrm{H}}(T) = a+bT^2$. The dashed lines in
figure~\ref{fg:001}(a) and (b) show such a fit for x=0.03 data,
with fitting parameters $\omega^{}_{\mathrm{H}} =
1.25\mathrm{cm}^{-1}$, $a = 1793\mathrm{cm}^{-1}$ and $b = 1.89
\times 10^{-2} \mathrm{cm}^{-1} \mathrm{K}^{-2}$. This
demonstrates that the transport revealed by Hall angle can be well
described by a simple Drude model and is similar to that of a
conventional Fermi liquid, even for the severely underdoped
$x=0.03$ sample.

%\begin{figure}
  % Requires \usepackage{graphicx}
  %\includegraphics[width=12cm]{fig002a.eps}\\
%  \centering
%  \input{fig002b1.tex}\\
%  \input{fig002b2.tex}
%  \includegraphics[width=8cm]{fig002a.eps}
%  \caption{Hall frequencies $\omega^{}_{\mathrm{H}}$ of the x=0.03,
%  0.10 and 0.15 samples.
%  }\label{fg:002}
%\end{figure}

Shown in figure~\ref{fg:003}(b) is the temperature dependence of
Hall frequencies $\omega^{}_{\mathrm{H}}$ for the three samples
obtained from Drude analysis:
$\omega^{}_{\mathrm{H}}=-\omega/\mathrm{Im}\
\theta_{\mathrm{H}}^{-1}$. A weak temperature dependence in
$\omega^{}_{\mathrm{H}}$ is observed in all three samples.
However,    drastic increase of $\omega^{}_{\mathrm{H}}$ is
observed when the hole doping level is reduced, confirming the
early observation~\cite{rigal01} of similar trend in a set of
underdoped YBCO samples. In the  case of YBCO, the analysis is
complicated by the existence of conducting CuO chains, whose
contribution to optical conductivity is difficult to  be reliably
subtracted.   For a Fermi-liquid the Hall frequency can be
expressed in terms of integrals of the Fermi velocity over the
Fermi surface as~\cite{cerne01}:
\begin{equation}\label{eq:003}
    \omega^{}_{\mathrm{H}}=\frac{eB}{\hbar c}\
    \frac{\displaystyle \oint_{\mathrm{FS}}\! \mathrm{d}S \bm{e}_z
    \cdot \left[\bm{v}(\bm{k}) \times
    \frac{\mathrm{d}}{\mathrm{d}k}\bm{v}(\bm{k})\right]}
    {\displaystyle
    \oint_{\mathrm{FS}}\!\mathrm{d}S|\bm{v}(\bm{k})|}.
\end{equation}
Therefore the observed strong increase in the Hall frequency  as
the hole doping is reduced from optimal doping suggests that the
Fermi surface topography changes significantly.  One possible
scenario is the gapping out of part of Fermi surface in underdoped
samples by the formation of a density wave state.  In case of
underdoped LSCO, there is evidence from ARPES ~\cite{yoshida01}
suggesting that, for severely underdoped $x=0.03$ sample, a
significant fraction of the large Fermi surface which is observed
in optimally doped samples is destroyed and the remaining small
portion is a small pocket near $(0.42\pi, 0.42\pi)$.  The observed
increase in Hall frequency in the density wave state is predicted
by a calculation~\cite{Tewari01} within the $d$-density
wave(DDW)model using the semiclassical Boltzmann theory in the
weak field limit~\cite{Chakravarty01}. However, there is an
alternative interpretation in terms of the effects of interactions
on the optical magneto-conductivity. Recent work by
Kontani~\cite{kontani02} based on the fluctuation exchange model
in which vertex corrections are included in the calculation of the
Hall conductivity also shows a Drude like response of the Hall
angle and an increase of $\omega^{}_{\mathrm{H}}$ by 70\% when
doping level is reduced from $x=0.20$ to $x=0.10$.  In choosing
between these two interpretations it is noteworthy that evidence
for a density wave gap has not been reported for the hole doped
cuprates.  By contrast Pr$_{2-x}$Ce$_x$CuO$_4$, an electron doped
cuprate, which exhibits a gap like feature in $\sigma_{xx}$,
displays a different $\sigma_{xy}$ response than the hole doped
cuprates in which features associated with density wave gap
excitations are seen~\cite{zimmers01,zimmers02}.

In summary, we studied the infrared Hall effect in a series of
LSCO samples with the hole doping level ranging from severely
underdoped ($x=0.03$ ) to optimally doped ($x=0.15$).  We found
that the doping and temperature dependence of Hall angle can be
well described with a simple Drude model.  A significant increase
of the Hall frequency is observed when the hole doping level is
reduced from optimal doping.  This finding can be ascribed to
either a drastic reduction of the volume of Fermi surface in the
underdoped sample due to the formation of a density wave state or
to the effects of vertex corrections on the Hall conductivity in a
strongly interacting electron system.  The absence of evidence for
a density wave gap in the $\sigma_{xx}$, and $\sigma_{xy}$ spectra
of LSCO indicates that the second interpreation is the correct
one.

The authors acknowledge the support of  NSF grant DMR-0303112.

\end{document}